\documentclass[showpacs,preprintnumbers,superscriptaddress]{revtex4}
\usepackage{epsfig}
\usepackage{bm}

\newcommand{\beq}{\begin{equation}}
\newcommand{\eeq}{\end{equation}}
\newcommand{\beqa}{\begin{eqnarray}}
\newcommand{\eeqa}{\end{eqnarray}}

\newcommand{\hepph}[1]{{\tt hep-ph/#1}}
\newcommand{\astroph}[1]{{\tt astro-ph/#1}}

\newcommand{\plb}[3]{Phys.\ Lett.\ B\ {\bf #1}, #3 (#2)}

\renewcommand{\apj}[3]{Astrophys.\ J.\ {\bf #1}, #3 (#2)}

\renewcommand{\prl}[3]{Phys.\ Rev.\ Lett. {\bf #1}, #3 (#2)}
\renewcommand{\prd}[3]{Phys.\ Rev.\ D\ {\bf #1}, #3 (#2)}

\begin{document}

\begin{flushright}
NSF-ITP-02-165\\ 
hep-ph/0211325
\end{flushright}

\title{Detection of Leptonic Dark Matter}

\author{E.~A.~Baltz}

\affiliation{Kavli Institute for Theoretical Physics, University of California
at Santa Barbara, Santa Barbara, CA 93106-4030}

\affiliation{Permanent address: ISCAP, Columbia Astrophysics Laboratory, 550 W
120th St., Mail Code 5247, New York, NY 10027 {\tt
eabaltz@physics.columbia.edu}}

\author{L.~Bergstr\"om}

\affiliation{Kavli Institute for Theoretical Physics, University of California
at Santa Barbara, Santa Barbara, CA 93106-4030}

\affiliation{Permanent address: Department of Physics, Stockholm University,
AlbaNova University Center, S-106 91 Stockholm, Sweden {\tt lbe@physto.se}}

\date{\today}

\begin{abstract}
Weakly interacting massive particles (WIMPs) are among the favored candidates
for cold dark matter in the universe. The phenomenology of supersymmetric WIMPs
has been quite developed during recent years. However, there are other
possibilities which have not been discussed as much. One example is a
right-handed massive neutrino, which has recently been proposed in the context
of a version of the Zee model for massive neutrinos.  This TeV-scale, leptonic
WIMP (or LIMP, for short) may at first sight appear to be essentially
undetectable. However, we point out that the radiatively induced annihilation
rate into leptons and photons is bound to be substantial, and provides a
conspicuous gamma-ray signature for annihilations in the galactic halo.  This
gives a window of opportunity for Air \v Cerenkov Telescopes with ability to
observe the galactic center, such as the HESS and CANGAROO arrays, and also for
the GLAST space telescope. In addition, the contribution to the positron cosmic
ray flux is in principle detectable, but this would require very strong local
density enhancements in the dark matter halo distribution.
\end{abstract}


\maketitle

\section{Introduction}\label{sec:intro}

Recently, Krauss, Nasri and Trodden (\cite{KNT}, KNT in the following) proposed
an interesting model, where a right-handed neutrino of mass on the order of a
few TeV plays a crucial role in giving mass to the otherwise massless standard
model neutrinos through a high-order loop mechanism. This is a version of the
Zee model \cite{zee}, which has been quite successful is reproducing the
observed mass and mixing pattern of solar and atmospheric neutrinos (see, e.g.,
\cite{jarlskog}). The particle content of the Zee model is given by two Higgs
doublets $\Phi_1$ and $\Phi_2$, and a charged field $S$ which transforms as a
singlet under $SU(2)$, with Lagrangian
\begin{equation}
{\cal L}_{\rm Zee} = f_{\alpha\beta}L_{\alpha}^TCi\tau_2L_{\beta}S^{+} + \mu
\Phi_1^Ti\tau_2\Phi_2 S^{-} + {\rm h.c.}
\end{equation}

KNT consider a variant where neutrino masses appear only at the three loop
level. To achieve this they supplement the SM fields with two charged singlet
scalars $S_1$ and $S_2$ and one right handed neutrino $N_R$.  Lepton number is
broken explicitly by including a Majorana mass term for the right-handed
neutrino, and imposing a discrete $Z_2$ symmetry under which the SM fields and
$S_1$ are singlets but $S_2$ and $N_R$ transform as
\begin{equation}
Z_2: \{S_2, N_R\} \longrightarrow \{-S_2, -N_R\} \ ,
\end{equation}
forbidding Dirac masses for the neutrinos.  This gives the Lagrangian
\begin{eqnarray}
{\cal L}_{\rm KNT}& = & f_{\alpha\beta}L_{\alpha}^TCi\tau_2L_{\beta}S_1^{+} +
g_{\alpha}N_RS_2^{+}l_{{\alpha}_R} \nonumber \\
&& + M_RN_R^TCN_R + V(S_1,S_2)
+ {\rm h.c.} \ ,
\end{eqnarray}
in which the potential $V(S_1,S_2)$ contains a $(S_1S_2^{*})^2$ coupling.  It
is assumed a mild hierarchy of masses $M_R<M_{S_1}<M_{S_2} \sim $ TeV and that
the Yukawa couplings $f_{\alpha\beta}$, $g_{\alpha}$ are of order unity, making
$N_R$ stable in view of the discrete symmetry.  Left-handed Majorana neutrino
masses are induced at three-loop order.  For $M_{S_2} \sim$ TeV, KNT find an
effective dimension-five effective mass scale of $\Lambda > 10^9$ GeV, giving
neutrino masses at the $0.1$ eV scale without involving fundamental mass scales
significantly larger than a TeV.

\section{Two-body tree-level annihilation rates}\label{sec:treelevel}

The discrete symmetry and the fact that $N_R$ is lighter than the charged
scalars means that $N_R$ becomes stable and therefore a natural dark matter
candidate. Through $S_2$ exchange it coupled to charged leptons in the early
universe strongly enough to give the correct relic density, but extremely
weakly today.  Current direct dark matter detectors employ scattering on
nucleons, and will not be sensitive to the available leptonic interactions.
One could imagine techniques based on atom level transitions \cite{spergel},
but a simple estimate shows that the rate is typically less than one event per
ton per year, thus several tons of active detector material would be needed --
a very difficult task given that a gaseous phase detector will probably be
necessary.

Similarly, indirect detection through neutrinos from the Earth or the Sun will
not be possible, since the cross section for capture is negligibly small.

We therefore turn to indirect detection through annihilation in the galactic
dark matter halo.  First, it may be useful to review the thermal production
mechanism in the early universe. Since the $N_R$ is a Majorana particle, its
annihilation into a lepton pair shows the usual helicity suppression $\sigma
v\propto m_\ell^2$ for the S-wave. Repeating the original calculation of
Goldberg for photinos \cite{goldberg}, the cross section can be written
\begin{equation}
\sigma v\left(N_RN_R\to \ell^+\ell^-\right)=
{g_\ell^4\over 8\pi m^4_N(1+f^2)^2}
\left[m^2_\ell+{2\over3}\left(\frac{1+f^4}{(1+f^2)^2}\right)\,m_N^2v^2
+...\right],
\label{eq:freezeout}
\end{equation}
where $m_\ell$ is the charged lepton mass, $m_S=fm_N$ the $S_2$ mass and $m_N$
the LIMP ($N_R$) mass.  Note that for $f$ not too much larger than unity, the
factor in square brackets is typically close to $0.5$.  This means that for
$m_N$ in the TeV range, the P-wave cross section (the term proportional to
$v^2$ in Eq.~(\ref{eq:freezeout})) will determine the freeze--out temperature
$T_f$ and therefore the relic abundance \cite{KNT}. As usual for a WIMP, one
finds $T_f/m_N\sim 1/20$, and so $\langle v^2\rangle_f=6T_f/m_N\sim 0.3$.  The
requirement that the LIMP be the dark matter particle, with relic abundance
$\Omega_N h^2\approx 0.1$, as indicated from a joint analysis of cosmic
microwave background and large scale structure data (see, e.g.,
\cite{cosmparams}), then fixes the cross section at freeze--out to be
\cite{jkg,lbreview}
\begin{equation}
\sum_{\ell=e,\mu,\tau}
\langle\sigma v\rangle_f\approx {\sum_\ell g^4_\ell\over 80\pi m_N^2
(1+f^2)^2}\approx 3\times 10^{-26}\ {\rm cm}^3\;{\rm s}^{-1}.
\end{equation}
For a given $N_R$ of mass $m_N$, this fixes the normalization of the
combination $\sum_\ell g^4_\ell/m_N^4(1+f^2)^2$ which appears in many
annihilation formulas. We can then estimate the total annihilation rate into
lepton pairs $\ell^+\ell^-$ at rest (i.e.\ in S-wave) putting $v=0$ in
Eq.~(\ref{eq:freezeout}). We find
\begin{equation}
\sigma v\left(N_RN_R\to \ell^+\ell^-\right)_{v\sim 0}\approx
3\times10^{-25}\;\left(\frac{g_\ell^4m^2_\ell}
{\sum_\ell g_\ell^4m^2_N}\right)\ {\rm cm}^3\;{\rm s}^{-1}=
10^{-25}\;\left(\frac{m^2_\ell}{m^2_N}\right)\ {\rm cm}^3\;{\rm s}^{-1}\;
({\rm flavor\ universal}).
\label{eq:rest}
\end{equation}
In the remainder of this paper we will assume flavor universality, namely
$g_e=g_\mu=g_\tau=g$.  For a TeV LIMP, this looks like a phenomenally small
annihilation rate for the $e^+e^-$ and $\mu^+\mu^-$ channels. Even for
$\tau^+\tau^-$, the helicity suppression is of the order of $10^{-6}$ (at that
level, the P-wave term in the galactic halo starts to contribute since typical
galactic velocities correspond to $v\sim 10^{-3}$). One would therefore be
tempted to conclude that the LIMP, despite being a good dark matter candidate,
has little chance to be detected in any direct or indirect detection
experiment.  We note that in order that the couplings $g_\ell$ not be much
larger than unity, for nearly degenerate $N_R$ and $S_2$, $m_N$ cannot be too
much heavier than 1 TeV, but could be significantly lighter.

\section{The detectability of $N_R$ through gamma-rays}\label{sec:contgammas}

The small annihilation rate, unlike the small scattering rate (which is caused
by the leptonic nature of the candidate), is due to the Majorana nature of
$N_R$ which implies the absence of large S-wave two-body annihilation at tree
level \cite{goldberg}. Any higher order process which does not have this
helicity suppression will have a chance to dominate the annihilation rate
completely. This is precisely the kind of interesting situation investigated
many years ago for the case of a pure photino coupling through light selectrons
to electrons, positrons and photons \cite{lbe89}. (That particular candidate is
now less plausible in view of results from the LEP accelerator.)  In
particular, we can immediately take over the results of \cite{lbe89} to write
the differential cross section for the radiative, non-helicity suppressed
process $N_RN_R\to \ell^+\ell^-\gamma$ as (here the annihilation rate $\Gamma=
\sigma v/2$)
\begin{equation}
{d\Gamma\over dE_\gamma dE_\ell}
\left(N_RN_R\rightarrow\ell^+\ell^-\gamma\right)=
3\times 10^{-25}\,\frac{g_\ell^4}
{\sum_\ell g_\ell^4}\,{\alpha(m^2_N+m^2_S)^2\over
\pi m_N}\,F(E_\gamma,E_\ell) \ {\rm cm}^3\;{\rm s}^{-1}\;{\rm
GeV}^{-2},\label{eq:result}
\end{equation}
where we have neglected the lepton masses (a good approximation also for the
$\tau^\pm$ for TeV LIMPs), and where
\begin{equation}
F(E_\gamma,E_\ell)={(m_N-E_\gamma)(2m^2_N-4m_NE_\ell-2m_NE_\gamma+2E^2_\ell+
2E_\ell E_\gamma+E^2_\gamma)
\over(3m^2_N-2m_NE_\ell-2m_NE_\gamma+m^2_S)^2(m^2_N-2m_NE_\ell-m^2_S)^2}.
\end{equation}

Note that the total rate for gamma rays of this process is fixed by the cross
section giving the relic density - it is independent of the values of the
individual $g_\ell$ in this massless limit.  The strength of this process is
seen to be of the order of $\alpha/\pi$ times the annihilation rate at
freeze-out, which is orders of magnitude larger than the helicity-suppressed
two-body S-wave annihilation rate.  This will mean, as we shall see, that there
is a hope of detecting the LIMP in gamma-rays, and perhaps also in an anomalous
positron component of the cosmic rays. On the other hand, since all three
charged SM leptons are too light to decay into baryons, indirect detection
through antiprotons (or antideuterons) is not expected for the LIMP.  A nice
feature of the result (Eq.~\ref{eq:result}) is its model independence.  For any
dark matter candidate of this type (a Majorana particle coupling mainly to
leptons through charged scalar exchange) we expect this strength of the
radiative annihilation signal.  This is to be contrasted with the situation in
the MSSM, where the predicted gamma-ray signal for a SUSY WIMP of any given
mass depends strongly on a large number of additional supersymmetric
parameters.  However, there are astrophysical uncertainties which make the
expected absolute fluxes difficult to estimate despite this robustness of the
particle physics properties.

By integrating Eq.~\ref{eq:result} over $E_\ell$ (with lower and upper
integration limits $m_N-E_\gamma$ and $m_N$, respectively) we get the
differential photon spectrum. Similarly, by integrating over $E_\gamma$ between
$m_N-E_\ell$ and $m_N$ the differential lepton spectrum (and the identical
anti-lepton spectrum) is obtained. Apart from the overall normalization, which
is fixed by the relic density and contains a factor $(m_N/m_S)^4$ strongly
favoring a scenario with at most a mild hierarchy of these masses, the shapes
of these distributions depend only on the scaled energies $E/m_N$ and the ratio
$m_N/m_S$.

\begin{figure}[!htb]
\begin{center}
\epsfig{file=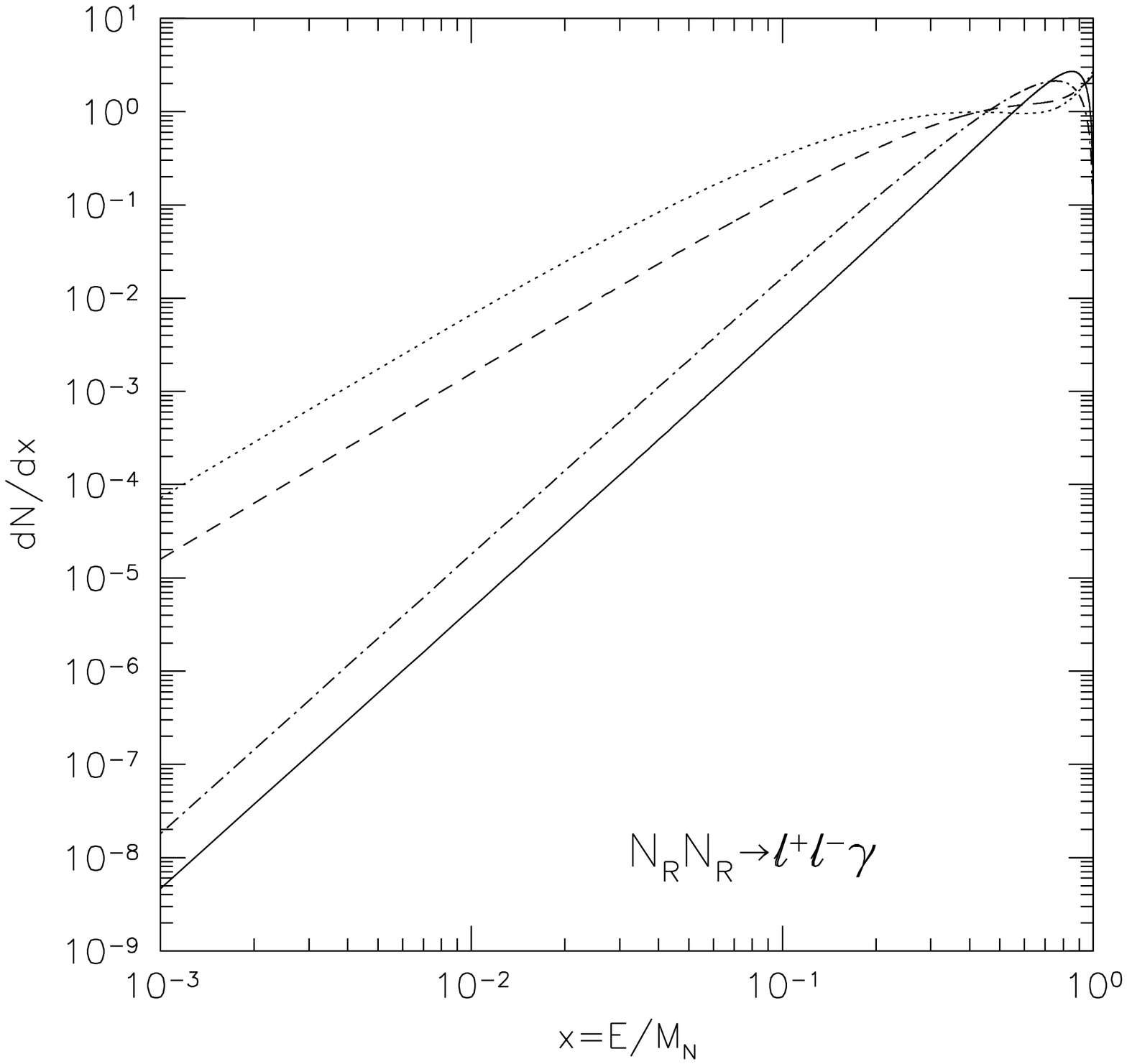,width=0.45\textwidth}
\epsfig{file=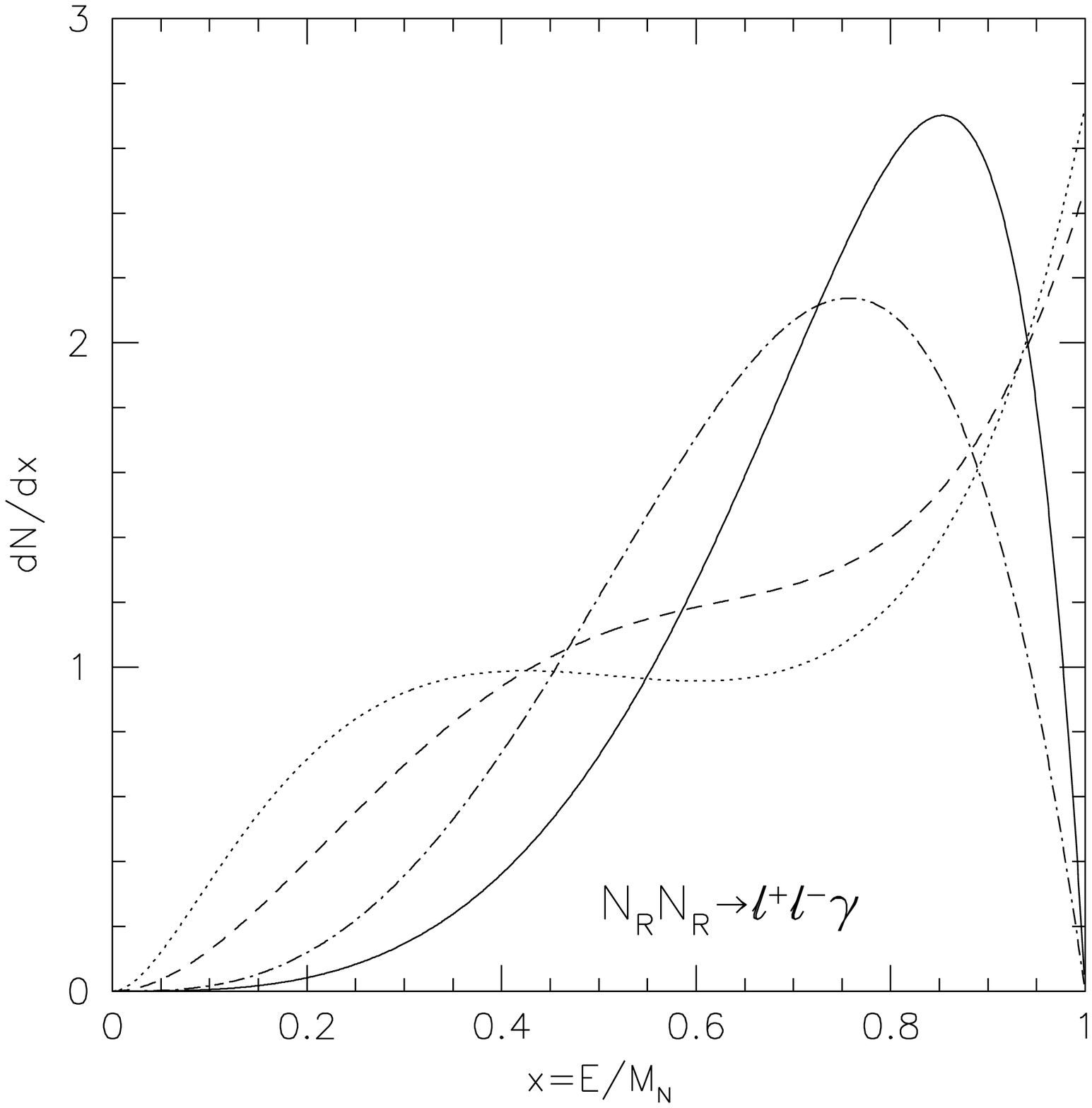,width=0.45\textwidth}
\end{center}
\caption{(a) The differential photon spectrum for the process $N_RN_R\to
\ell^+\ell^-\gamma$, normalized to unity, for $m_N/m_S=0.8$ (solid line) and
$m_N/m_S=0.2$ (dash-dotted line), as well as the lepton (or the identical
anti-lepton) spectrum for the same mass ratios (dotted line and dashed line,
respectively). (b) Same as in (a) but plotted using a linear scale on both
axes.}
\label{fig:fig1}
\end{figure}

Since all leptons can be treated as massless, the distributions are identical
for $e^\pm\gamma$, $\mu^\pm\gamma$ and $\tau^\pm\gamma$, so from the point of
view of this gamma-ray signature, it is of no importance whether the $N_R$
couples universally to leptons or not.  There are in principle a significant
number of hard photons in the tau decay chain from pion decays, but we have
checked that the direct photons from $\ell^\pm\gamma$ strongly dominate in all
cases we illustrate.  For, e.g., the positron signal there may be differences,
since muons and taus give positrons with softer spectra in their decays, as
discussed in the next Section.  In Fig.~\ref{fig:fig1} (a) and (b) we show
examples of these distributions for $m_N/m_S=0.8$ and $0.2$, plotted on
logarithmic and linear scales, respectively. As can be seen, both the gamma and
the lepton spectra are exceptionally hard, peaking near the maximum energy
(which in both cases is equal to $m_N$). This is optimal from the point of view
of detection, as most conceivable sources of background tend to give soft
spectra, rapidly falling with energy.

The obvious place to search for the gamma-ray signal is in the direction of the
galactic center, where the dark matter distribution is expected to be strongly
enhanced. N-body simulations indicate that cold dark matter, of which the LIMP
is an example, is even likely to form a density cusp $\rho(r)\sim 1/r$ near the
center (the so-called Navarro-Frenk-White or NFW profile \cite{NFW}). The
observational situation concerning rotation curves and the distribution of dark
matter in spiral galaxies, including our own, is however far from clear at the
moment, and the same is true for the N-body calculations where different groups
still seem get different results \cite{moore_review,primack}.  Even more
unclear is the interesting possibility of the massive black hole transforming
the cusp into a very sharp spike with dramatically enhanced density
\cite{gondolo_silk,ullio,merritt}.

The best we can do at the moment is to assume an NFW profile, keeping in mind
that the predicted rates can be either much lower (in the case of the absence
of both a cusp and a spike) or much higher (in the case of a steeper cusp, as
some N-body simulations indicate, or a spike, or a clumpy halo \cite{clumpy1}
as also indicated by simulations \cite{clumpy2}).  Here we note the importance
of the clumpiness of the halo: since the annihilation rate is proportional to
the square of the density $\rho$, it is enhanced over the prediction for a
smooth halo by a clumpiness parameter
$C=\langle\rho^2\rangle/\langle\rho\rangle^2\ge1$.

The rate of gamma-rays from a given halo profile has been elaborated in
\cite{BUB}. Combining the results of that work with those given here, we
predict a gamma-ray flux from the galactic center (treated as a point source,
so the units are in ${\rm cm}^{-2}\;{\rm s}^{-1}\;{\rm GeV}^{-1}$)
\begin{equation}
{d\Phi_\gamma\over dE_\gamma}=1.3\times 10^{-11}\left({1\ {\rm TeV}\over
m_N}\right)^2\left({\langle J\rangle_{\Delta\Omega} \Delta\Omega\over 100}
\right){(m^2_N+m^2_S)^2\over m_N}\int_{m_N-E_\gamma}^{m_N}dE_\ell
F(E_\gamma,E_\ell) , \label{eq:contgammas}
\end{equation}
where the line-of-sight integral $J$ is defined in \cite{BUB}, and $\langle
J\rangle_{\Delta\Omega}$ is the average of this line-of-sight integral over the
acceptance $\Delta\Omega$ of a gamma-ray telescope; for a NFW profile, the
maximal value of the product $\langle J\rangle_{\Delta\Omega}\Delta\Omega \sim
100$ for $\Delta\Omega\sim 10^{-3}$ \cite{BUB} (similar, or larger, values
would be obtained for the Moore profile \cite{moore} $\rho(r)\sim 1/r^{1.5}$).
The background is difficult to estimate (in fact, much of the presently
perceived diffuse ``background'' may turn out to be part of the kind of signal
discussed here). For the sake of the argument, we will take the simple estimate
given in \cite{BUB}, extrapolated from the flux measured by EGRET towards the
galactic center:
\begin{equation}
{\left(d\Phi^{bkg}_\gamma\over dE_\gamma d\Omega\right)_{g.c.}}= 4.8\times
10^{-13}\left({1\ {\rm TeV}\over E_\gamma}\right)^{2.7}\ {\rm cm}^{-2}\;{\rm
s}^{-1}\;{\rm GeV}^{-1}\;{\rm sr}^{-1}.\label{eq:bkggammas}
\end{equation}

An inevitable consequence of the LIMP coupling to a charged scalar is that the
loop-induced process $N_RN_R\to\gamma\gamma$ should also occur \cite{lbe89},
with cross section
\begin{equation}
\sigma v\left(N_RN_R\to \gamma\gamma\right)\approx {\left(\sum_\ell
g^2_\ell\right)^2\over \sum_\ell g^4_\ell}\times 4\times 10^{-30} \ {\rm
cm}^3\;{\rm s}^{-1}.
\end{equation}
Note that since all leptons contribute coherently to the loop amplitude, the
rate is more favorable for a universal lepton coupling. In the case
$g_e=g_\mu=g_\tau=g$, the value for $\sigma v$ becomes $1.2\times
10^{-29}\;{\rm cm^3\;s^{-1}}$ which is quite sizable. Note that this result is
independent of the $N_R$ and $S_2$ masses \cite{lbe89}. (One may also note that
since $S_2$ is an SU(2) singlet, there is no corresponding $Z\gamma$ final
state.)

The line flux then becomes 
\begin{equation}
\Phi_{\gamma\gamma}^{g.c.}(E_\gamma=m_N)\approx 1.2\times 10^{-13}\left({1\
{\rm TeV}\over m_N}\right)^{2}\ {\rm cm}^{-2}\;{\rm
s}^{-1}.\label{eq:linegammas}
\end{equation}
This would appear as a sharp spike (with relative width of the order of the
Doppler broadening due to LIMP motion, i.e. of the order of $10^{-3}$), in an
instrument with perfect energy resolution.  However, ACTs have an energy
resolution of the order of $5 - 10$\% at best.  If we assume 5\%, the gamma
spike for a TeV LIMP spreads out over 50 GeV, a differential rate near a TeV
similar to the radiative processes discussed above.

For smaller LIMP masses of the order of 100 GeV, one gets into the mass range
where the GLAST gamma-ray telescope \cite{glast} will be operative (it will
detect gamma-rays up to 300 GeV).  The energy resolution for GLAST will depend
on the angle of incidence and may reach a few percent, making it possible for
the line to stand out against background.  The detection of such a line would
eliminate all possible confusion with plausible astrophysical backgrounds.

\begin{figure}[!htb]
\begin{center}
\epsfig{file=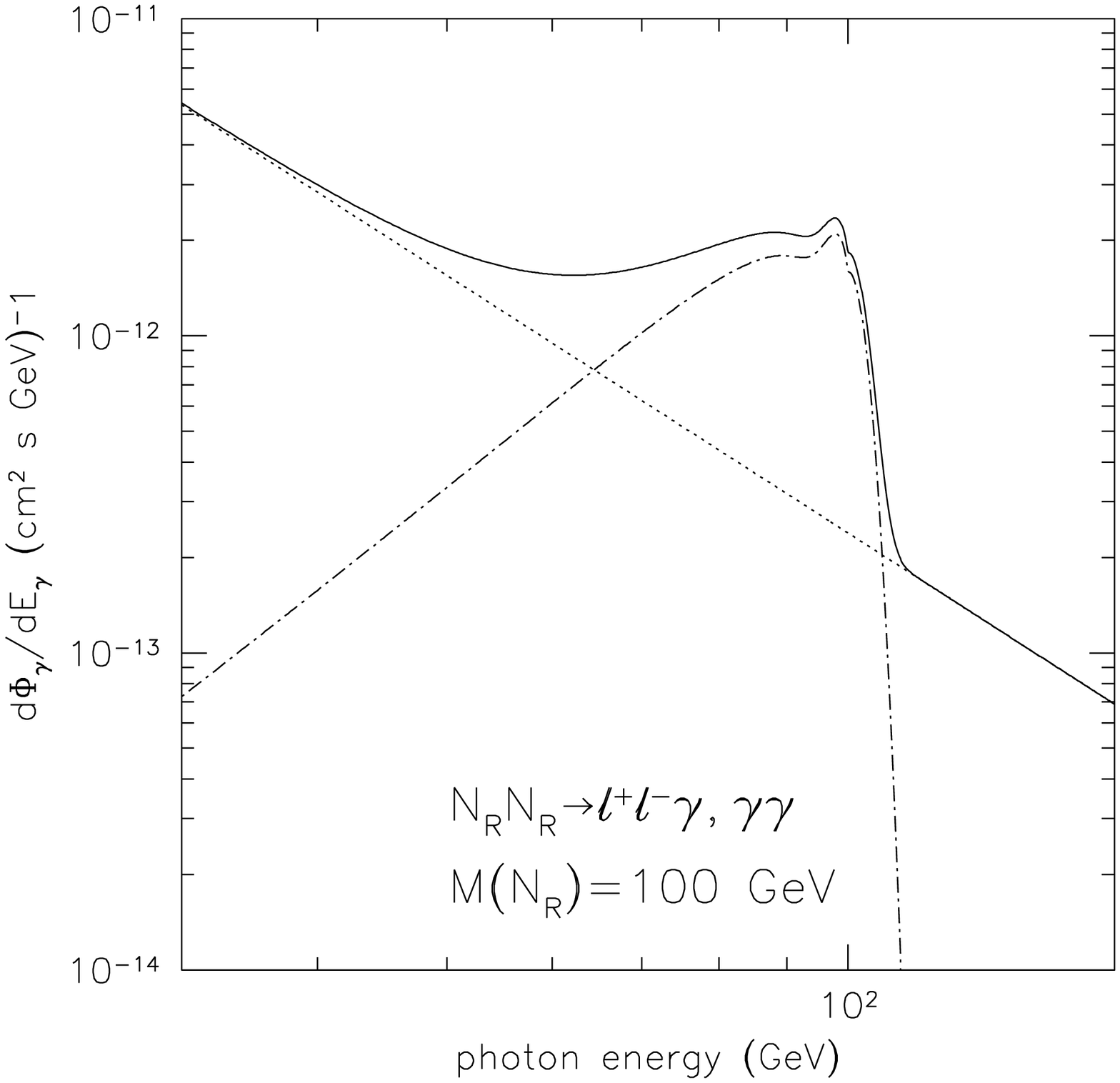,width=0.32\textwidth}
\epsfig{file=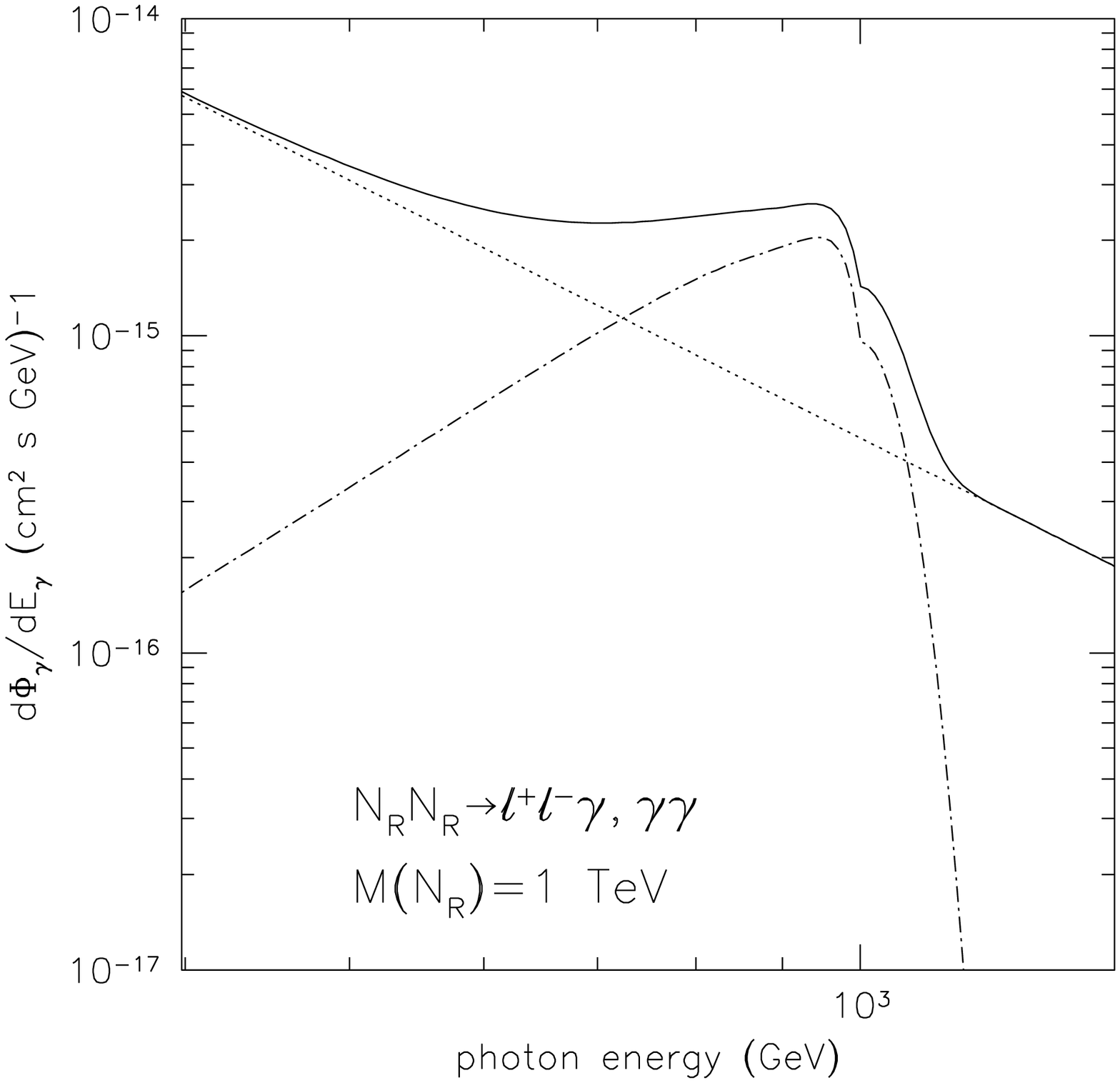,width=0.32\textwidth}
\epsfig{file=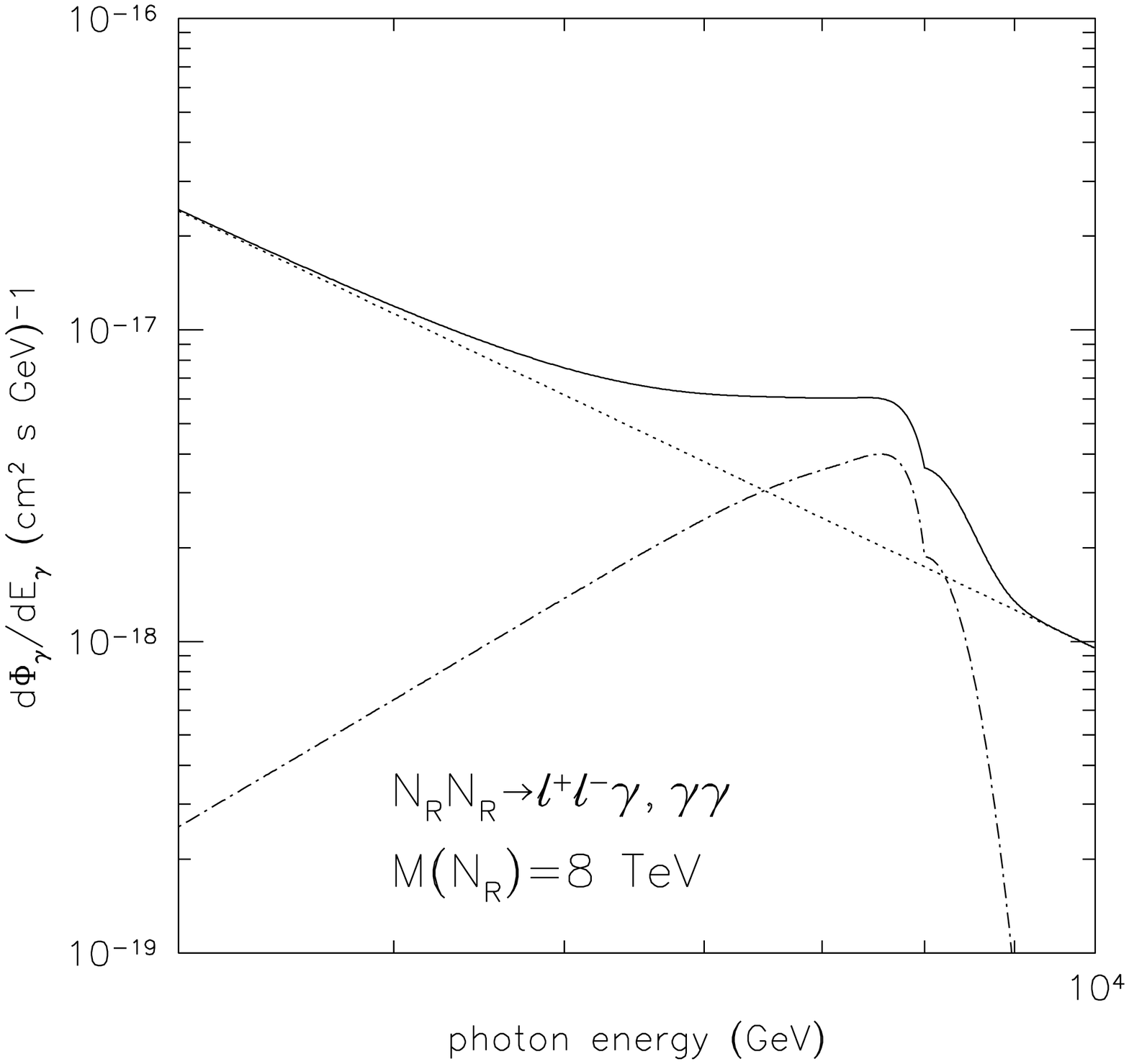,width=0.32\textwidth}
\end{center}
\caption{(a) The total gamma-ray flux expected from a $\Delta\Omega=10^{-3}$ sr
cone around the galactic center (solid line). The flux is composed by a
power-law background extrapolated from EGRET data (dotted line) and a 100 GeV
LIMP annihilating with a cusped (NFW) density profile through a 110 GeV scalar
$S_2$, giving both a continuous spectrum and a 2$\gamma$ line.  An energy
resolution of 3\% has been assumed for the line signal. (b) Same as (a) for a 1
TeV LIMP, $m_{S_2}= 1.1$ TeV. Here the line has been smeared by an assumed
energy resolution of 5\%. (c) Same as (b) for an 8 TeV LIMP, $m_{S_2}=8.8$
TeV.}
\label{fig:fig2}
\end{figure}

In Fig.~\ref{fig:fig2} (a) we show the flux predicted from
Eq.~(\ref{eq:contgammas}) and Eq.~(\ref{eq:linegammas}) together with the
background estimate Eq.~(\ref{eq:bkggammas}) for a 100 GeV LIMP, and 110 GeV
$S_2$, NFW profile and $\Delta\Omega= 10^{-3}$.  For this energy range, we have
used an energy resolution of 3\%.  It may be difficult to push the $N_R$ mass
much below 100 GeV without fine tuning the parameters of the KNT model (and the
$S_2$ mass is also bounded by LEP results to be larger than around 100 GeV).
We note that with the $S_2$ (and thus $S_1$) only 10\% higher in mass than the
$N_R$, a careful analysis should really take coannihilations with $S_2$ into
account when estimating the relic density, and thereby fixing the interaction
strength.  In analogy to SUSY models where R-parity stabilizes the lightest
superpartner, in this model the $Z_2$ symmetry stabilizes the lightest odd
state; a complete calculation would include all odd states, in this case $N_R$
and $S_2$.  However, coannihilations are unlikely to change the relic density
by more than a factor of a few, a small correction compared with the very large
astrophysical uncertainties related to the radial distribution of dark matter
near the galactic center.

The natural mass range for the LIMP is around 1 TeV, where GLAST runs out of
sensitivity but where ground-based arrays of Air \v Cerenkov Telescopes with
large collecting area can detect a signal. Indeed, there are already such
arrays of telescopes planned or in operation such as CANGAROO \cite{cangaroo},
HESS \cite{hess}, VERITAS \cite{veritas} and MAGIC \cite{magic}. In particular,
CANGAROO and HESS are well located to observe the galactic center for a sizable
fraction of their observing time. As can be seen from the figure, the signal
with these assumptions would stand out from the gamma-ray background. (We do
not enter here into the more technical issue of rejecting other types of
background, such as from hadrons and electrons, where there is a steady
improvement in the techniques employed.)

In Fig.~\ref{fig:fig2} (b) and (c), the curves are shown for a LIMP of mass
$m_N=1$ TeV, $m_{S_2}=1.1$ TeV, and $m_N=8$ TeV, $m_{S_2}=8.8$ TeV,
respectively.  These should be clearly observable with a very conspicuous
``bump'' in the spectrum, for the halo parameters chosen. We note with interest
that preliminary results from the CANGAROO collaboration indeed show an excess
flux of TeV gamma-rays from the galactic center \cite{kashiwa}.  The absolute
flux level for this possible signal seems higher than that predicted in
Eq.~(\ref{eq:contgammas}), so an enhancement beyond that provided by the NFW
profile would then be indicated.

An interesting question is what would happen to the leptons. To be specific,
assume that there is lepton universality in the $N_R$ couplings. Then there
will be equal amounts (apart from very small lepton mass corrections) of
$\tau^\pm$, $\mu^\pm$ and $e^\pm$ leptons. The produced $\tau^\pm$ and
$\mu^\pm$ will decay into lower-energy electrons and positrons (see next
Section), and the primary high-energy electrons and positrons will quickly
radiate due to synchrotron and inverse Compton processes.  In fact, an analysis
by Bertone, Sigl and Silk \cite{bertone} points to the existence of a radio
signal that may be caused by just the kind of TeV-scale dark matter particles
discussed here, if the magnetic field near the galactic center is strong
enough. In any case, this would be a signal to search for, were a TeV gamma-ray
excess to be confirmed.

\section{Predictions for cosmic-ray positrons}\label{sec:positrons}

In the previous section the annihilation process
$N_RN_R\rightarrow\ell^+\ell^-\gamma$ was discussed with the idea of detecting
the high energy photons produced.  We can also make a prediction for the cosmic
ray positron flux from this process.  As shown below, the positron flux is
quite low, and not likely to be detectable unless the clumpiness factor of the
galactic halo is quite large.

Positrons might come from any of the three leptonic channels.  In the direct
process $N_RN_R\rightarrow e^+e^-\gamma$ the positrons are quite energetic.
The energy is downgraded somewhat in the muon channel, since some of the energy
goes to neutrinos: $N_RN_R\rightarrow \mu^+\mu^-\gamma\rightarrow
e^+e^-\nu_e\overline{\nu}_e\nu_\mu\overline{\nu}_\mu\gamma$.  The spectrum is
softest for the tau channel $N_RN_R\rightarrow\tau^+\tau^-\gamma$, since there
are many hadronic modes with longer decay chains yielding positrons (and
electrons) in the end.  We also include the simpler processes
$N_RN_R\rightarrow\ell^+\ell^-$.  For electrons this is negligible unless $M_N$
is very small and $M_S\gg M_N$ is much larger.  However, for muons and
especially taus, the simple process can be important as the helicity
suppression is less severe.  All of this can be accounted for with standard
tools; in particular we have used results from the PYTHIA event generator
\cite{pythia} as tabulated in DarkSUSY \cite{DarkSUSY} for the positron spectra
from the muon and tau decay chains.  Folding in the annihilation cross section,
we can then compute the volume production rate of positrons in the Galactic
halo ($d\Gamma/dE$) in units of cm$^3$ s$^{-1}$ GeV$^{-1}$.

The propagation of positrons in the Galactic environment is complex.  The
gyroradii of charged particles in the tangled magnetic fields are quite small,
and the motions can be modeled as a diffusion process.  Furthermore, positrons
lose energy rapidly to both synchrotron radiation, and in addition to inverse
Compton scattering with the cosmic microwave background and with diffuse
starlight.  We use the propagation model in Ref.~\cite{be} to calculate the
observed flux at the Earth.  The model includes diffusion in an infinite slab,
a reasonable approximation given the fact that detectable positrons are
produced within a few kpc, thus an outer radial boundary is unimportant.  The
energy losses to synchrotron and inverse Compton processes are included.  This
model is in rough agreement with earlier work \cite{kamturner}, though the
inclusion of inverse Compton scattering from starlight doubles the energy loss
rate.  More sophisticated models give similar results \cite{ms99}.  We note
here that the question of the halo profile is much less important in
calculating the positron flux, as their effective range is only a few kpc.
Thus, it is the less uncertain local dark matter density that is important for
the positron flux.

\begin{figure}[!htb]
\begin{center}
\epsfig{file=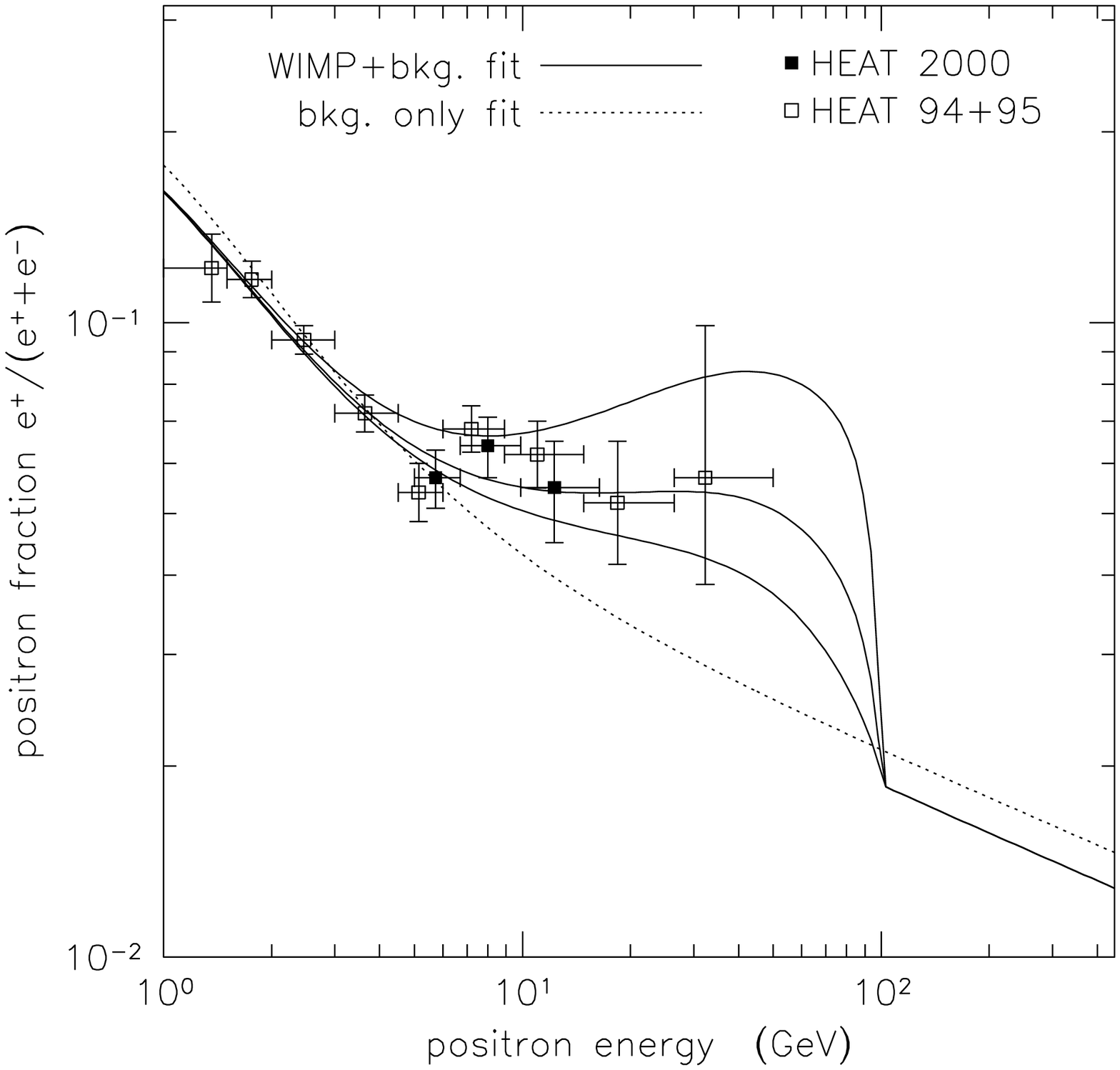,width=0.45\textwidth}
\epsfig{file=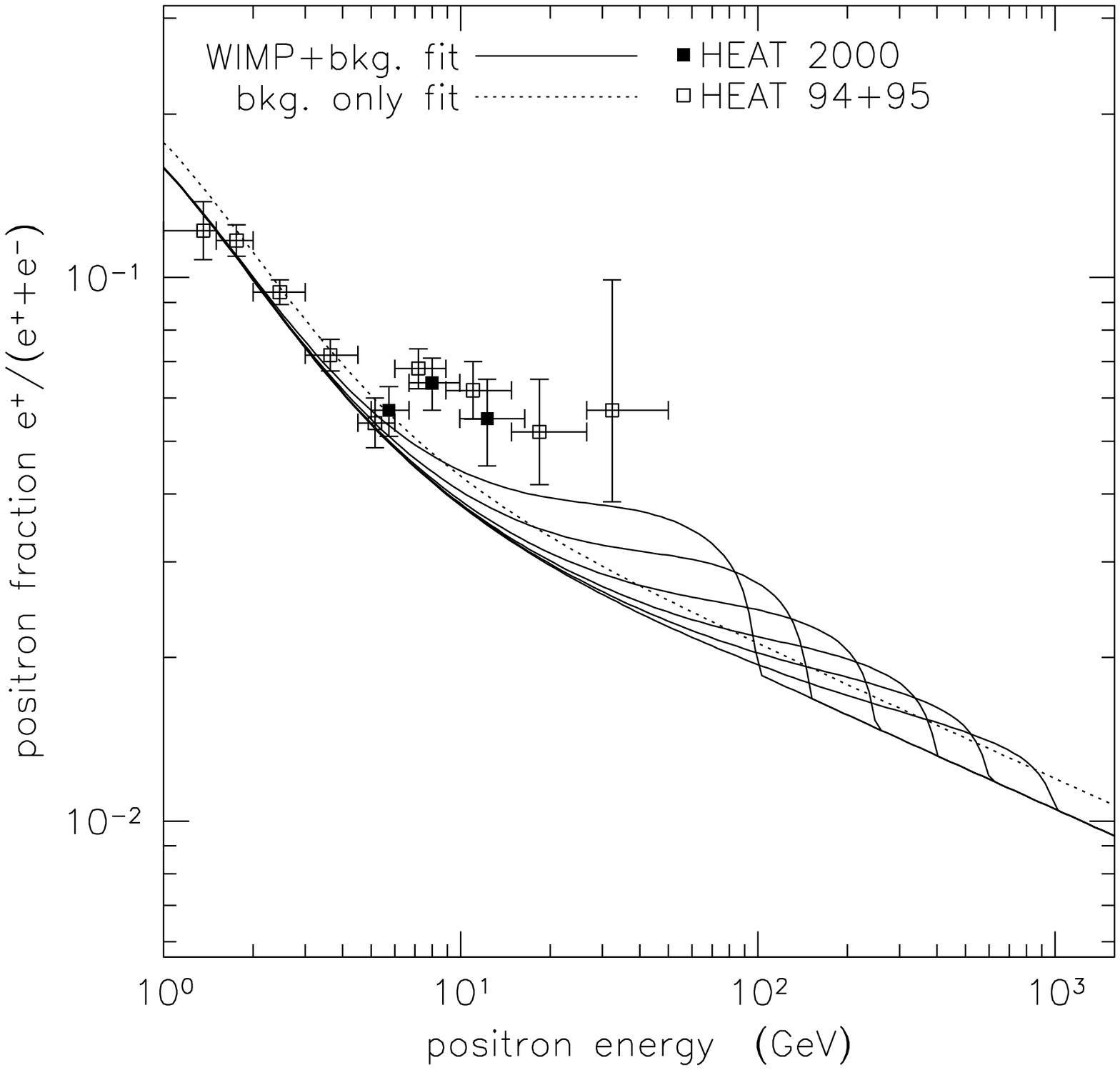,width=0.45\textwidth}
\end{center}
\caption{Cosmic ray positron fraction for LIMP models.  The data from the two
HEAT experiments in illustrated, along with the expected value (dotted line).
(a) Variation of $M_S$ with values of (top to bottom) 120, 150, and 200 GeV,
with $M_N$ fixed at 100 GeV.  The boost factor is fixed at $2\times10^4$.  (b)
Variation in $M_N$, with values of 100, 150, 250, 400, 600, and 1000 GeV.  In
all cases $M_S=1.2 M_N$.  Here a boost factor of $4\times10^3$ is used.}
\label{fig:fig3}
\end{figure}

The HEAT collaboration has in fact reported an excess of positrons above about
10 GeV in the cosmic ray positron fraction $f=e^+/(e^++e^-)$ \cite{heatfrac},
later confirmed in a different instrument by HEAT-pbar \cite{newdata}.  We now
discuss the possibility that this excess might be due to annihilating LIMPs.
The fluxes of positrons are typically too small by a factor of $10^4$ if a
smooth halo is assumed.  This might seem hopeless, but the clumpiness parameter
could be this large, boosting the signal accordingly.  In order to produce an
excess at 10 GeV, lighter LIMPs are preferred.  Assuming that the LIMP must be
heavier than 100 GeV to escape discovery (actually, it is the heavier charged
scalars $S_1$ and $S_2$ which would be seen at e.g.\ LEP), we illustrate some
LIMP models that give an acceptable positron flux to explain the HEAT results
in Fig.~\ref{fig:fig3} (albeit with the very large clumpiness parameter).
Other explanations for this excess have been put forward elsewhere
\cite{be,kanepositron,befg}.  In the second panel of Fig.~\ref{fig:fig3} we
show some LIMP models with larger masses that do not do well in explaining the
current excess.  However, future experiments such as AMS \cite{AMS} and PAMELA
\cite{PAMELA} may be able to probe such models with higher accuracy, and we
estimate that boost factors a factor of five or ten smaller (of order $10^3$)
would still produce an interesting signal for such experiments.

\section{Discussion and Conclusions}\label{sec:conclusions}

We have discussed the possibility of astrophysically detecting a right--handed
neutrino cold dark matter candidate arising in a recently proposed modified Zee
model.  As the interactions are only leptonic, current elastic scattering
experiments are not sensitive to this particle.  Annihilations in the galactic
halo, with both two and three body final states, may provide detectable numbers
of high energy photons or positrons, though these predictions rely heavily on
the structure of the Galactic halo, both in core profile and in spectrum of
substructure.

\section*{Acknowledgments}

We would like to thank L.~Krauss and J.~Edsj\"o for interesting conversations.
We thank the organizers of CMB02 at the Kavli Institute for Theoretical
Physics, where this research was done.  This research was supported in part by
the NSF under grant PHY99-07949 at KITP.  The work of LB was also supported by
the Swedish Research Council (VR).

\end{document}